\documentclass[aps,pre,showpacs,epsfig]{revtex4}
\usepackage{graphics}
\begin {document}

\def\be{\begin{equation}}
\def\ee{\end{equation}}
\def\bea{\begin{eqnarray}}
\def\eea{\end{eqnarray}}
\def\beq{\begin{eqnarray*}}
\def\eeq{\end{eqnarray*}}
\def\ba{\begin{array}}
\def\ea{\end{array}}

\def\ds{\displaystyle}

\newcommand{\bfDel}{\mbox{\boldmath $\Delta$}}
\newcommand{\bfwp}{\mbox{\boldmath $\wp$}}
\newcommand{\bfro}{\mbox{\boldmath $\rho$}}
\newcommand{\bfd}{\mbox{\boldmath $\delta$}}
\newcommand{\bfxi}{\mbox{\boldmath $\xi$}}

\def\m{{\rm m}}
\def\M{{\rm M}}
\def\ch{{\rm ch}}

\def\rr{{\bf r}}
\def\RR{{\bf R}}

\def\k{{\rm k}}

\def\ae{\tilde{\chi}}

\title{Modeling DNA conformational transformations on the mesoscopic
scales}

\author{S. N. Volkov}
\affiliation{Bogolyubov Institute  for Theoretical Physics, Kiev
03143, Ukraine, e.mail:snvolkov@bitp.kiev.ua}

\date{\today}

\begin{abstract}

The approach for the description of the DNA conformational
transformations on the mesoscopic scales in the frame of the
double helix is presented. Due to consideration of the joint
motions of DNA structural elements along the conformational
pathways the models for different transformations may be
constructed in the unifying two-component form. One
component of the model is the degree of freedom of the
elastic rod and another component -- the effective coordinate
of the conformational transformation. The internal and
external model components are interrelated, as it is
characteristic for the DNA structure organization. It is
shown that the kinetic energy of the conformational
transformation of heterogeneous DNA may be put in
homogeneous form. In the frame of the developed approach the
static excitations of the DNA structure under the transitions between
the stable states are found for internal and external components.
The comparison of the data obtained with the experiment on intrinsic
DNA deformability shows good qualitative agreement. The
conclusion is made that the found excitations in the DNA
structure may be classificated as the static conformational
solitons.
\end{abstract}


\maketitle

\section{INTRODUCTION}

The study of the conformational transformations of DNA double
helix is the line of understanding the mechanisms of the living
systems functioning. Last decade particular attention attracts the
conformational transformations of DNA macromolecule, which include
the displacements of the structural elements in the frame of
double helix with relatively large deviations from equilibrium
positions. The amplitudes of these displacements are not small,
but on the other hand -- not so large to destroy the double helix
organization. Such DNA transformations in the frame of double
helix we will name as mesoscopic, leaving the term "large-scale"
for the displacements of the structural elements that can destroy
the helix. The DNA mesoscopic transformations show themselves in
the formation of intermediate $A$-$B$ conformations in the complex
of DNA TATA-box with proteins [1-4], the motifs of $A$-form in
ligand-bound DNA (see [5] and cited therein), the junctions of
$B$- and $A$-forms in DNA fragments bound to the protein or drug
[6-9], the bistable flexibility of A-tract [10-12], the base-pair
preopening in oligonucleotide duplex [13], the overstretching of
DNA helix under high tension [14,15], and in a number of unusual
DNA conformations [16]. In mentioned above conformation
transformations the displacements of DNA structural elements are
accompanied with the deformation of DNA macromolecule as a whole,
but without the destruction of the double helix.

The investigation of the mesoscopic transformations is of great
interest because these DNA rearrangements directly connect with
the regulation of the genes activity [1,17]. From a complex
structure of DNA macromolecule the experimental study of its
conformational mobility meets with some obstacles [18]. One
of the productive way to understand the conformational resources
of the double helix is the use the phenomenological approach to
modeling its conformational transformations. The exploring such
models as "helix-coil" [19-21] or "elastic rod" [22-26] speeds up
and extends the investigations of the conformational properties of
the nucleic acids, and allow to interpret a set of experimental
data on DNA melting, helix bending and torsion, loop formation and
higher order folding. The model approach has definite advantages
comparing to various all atoms computational methods. From the
large number of atoms in a macromolecule the use of the
computational approach is restricted by the fragments of DNA helix
not longer than 20 base pairs [27,28]. At the same time under
successful model construction the phenomenological approach allows
to describe the structural transformations in macromolecules of
the real size and with the account of surrounding factors.

For modeling DNA mesoscopic transformations it would be
advantageous to use an experience in the modeling of
large-amplitude mobility of the structural elements in DNA double
helix, and  the study of the possibility of the localized
excitations realization [29]. On this direction quite a number of
models have been constructed (see review [30], collected articles
[31] and book [32]). But not all of these studies achieved real
results because of the significant simplifications of the models
and the difficulties of tacking into account such principle
properties of DNA structure, as nucleotide content and sequence.

Recently, a set of the models for the reduced description of the
internal mobility of DNA structural elements on the large scales
has appeared [33-37]. The authors of this studies divide the atoms
of the double helix on the atomic groups -- the new "structural"
elements of the system, and try to find the potential field which
could describe the interactions between them. This is a very
complicated task which cannot be resolved for DNA as a whole.

In the present work the another way is proposed. In modeling DNA
conformational transformations on the mesoscopic scales the four-mass
model approach is used. This approach was developed earlier [38-40] for
the description of DNA double helix conformational vibrations. The
achieved agreement between theory and experiment in the description of the
DNA low-frequency spectra [39,40] show the accordance of the four-mass
model to the described mobility of the double helix. It would appear
reasonable that the displacements of DNA structural elements with small
amplitudes are the precedes of the mesoscopic and large amplitude
transformations. It is taken into account also that under double helix
transformations the structural elements move jointly along the
conformational pathways. Therefore, starting from the four-mass model for
DNA internal mobility the models for definite pathways of the double helix
transformations are constructed. Such an approach allows us to reduce
significantly the degrees of freedom of a system and to choose the
adequate potential functions for described force field.

Using the proposed approach the models for the description of DNA
conformation transition of $A$--$B$ type and double helix opening
are constructed. It is shown that different conformational
rearrangements of the structural elements in the double helix may
be described by the unifying two-component model. The accounting
of the interrelation between the model components is discussed as
an important part of the description of DNA transformations.
In the frame of the two-component model the static excitations
are found for two types of the conformational transformations of
the macromolecule: transition from the ground state to metastable
and transition between the equivalent states in the conditions of DNA
bistasbility (for example, $A$--$B$ equilibrium). The obtained results
are compared with with the experimental data on DNA deformability.

\section{MODEL CONSTRUCTION}

The DNA molecule consists of two strands which are formed by the
linked phosphate groups and deoxyribose (or sugar) rings. To the
sugar rings the nucleic bases are attached. The DNA bases have
different content of atoms, that is why the primary structure of a
macromolecule is heterogeneous and may be used for writing the
genetic information. In the natural conditions DNA strands form
the double helix -- DNA secondary structure, where the nucleic
bases of different strands are connected by hydrogen bonds in the
complementary pairs A$\cdot$T and G$\cdot$C.

The DNA rearrangements on non-small scales in the frame of the
double helix have a character of transitions from one stable state
to another [18]. Under conformational transformations the
disposition of the structural elements in double helix monomer
units changes, and the macromolecule as a whole suffers definite
deformation. The form of macromolecule deformation and the type of
the conformational state of the double helix are observed in the
experimental studies [1-16]. So, for modeling the mesoscopic
transformations of DNA double helix it is important that the
constructed model will include the components for the description
of the internal conformational transformation and the external
deformation of macromolecule. For construction the models of
definite DNA transformations let us consider the four-mass model
approach.

\subsection{Generalized four-mass model}

According to approach of four-mass model for the DNA monomer link [38-40]
the basic conformational degrees of freedom, which characterize the
mobility of the structural elements of a double helix, are the following.
First of all, the displacements of the nucleosides (nucleic bases together
with sugar rings) as rigid pendulums "suspended" to the backbone (Fig.
1a). Secondly, intranucleoside transformations: the displacements of
nucleic bases under the changes in the sugar rings conformation. The third
type is the displacements of a nucleotide as a whole (nucleic base + sugar
+ phosphate group).

For the description of DNA structural dynamics in the approach
[38-40] two masses of the pendulum-nucleosides ($\m$) and two
masses of the backbone atom groups ($\m_0$) (Fig.~1) were used.
The masses of nucleosides have the different values according to
the kind of the nucleic base. The backbone masses $\m_0$ include
the group of atoms: PO$_2$ + 2O + C$_{5'}$ and are identical along
the chain. The necessary structure parameters of the model, such
as the length of the physical pendulum-nucleosides -- $l_0$, and
the angles of their position in the pair plane -- $\Theta_0$
(Fig.~1) calculate from the known double helix conformation. For
DNA B-form the calculated parameters of the model are shown in the
Table~I. As seen, four-mass model takes into account an essential
property of the DNA macromolecule -- heterogeneity of its monomer
content. Really, the model masses and geometrical parameters of
the pendulum-nucleosides depend on the nucleic bases contained in
its constitution.

Let us consider the possibility of using the four-mass model for
description of the mesoscopic transformations of DNA double helix. We will
write the expression for the energy of generalized four-mass model without
envisaging the smallness of displacements:
\be E = {1\over2} \sum_{i,n} \left \{
\m_0\dot{\RR}^2_{i,n}+\m_i\dot{\rr}^2_{i,n} +U({\rr}_{i,n};\RR_{i,n})
\right\}.  \\[0.3cm]
\ee
In the expression (1) the summation is made over all monomer links ($n$)
of the double chain ($i=1,2$). The radius-vectors $\RR_{i,n}(t)$ and
$\rr_{i,n}(t)$ describe the displacements of the mass centers of the
backbone groups and nucleosides, respectively, in each chain. The
$\dot{\RR}$ and $\dot{\rr}$ denote the time derivatives. The function $U$
is the potential energy of conformational transformations of a system.

Taking into account the structural organization of a macromolecule, the
potential energy of the n-th monomer link would be presented as:
\be U(n)=U^{(1)}(n)+U^{(2)}(n,n-1)+U^{(3)}(n;n \pm 1). \ee
The first term in the sum (2) describes the energy related to the internal
rearrangements in n-th monomer of the macromolecule as such. In the
four-mass model this energy may be written as:
\bea U^{(1)}&=&U_{1}^{(1)}[\RR_1-\rr_1]+U_{2}^{(1)}[\RR_2-\rr_2] +
\nonumber \\[0.3cm]
&+&U_{3}^{(1)}[\rr_1-\rr_2] +U_{4}^{(1)}[\RR_1-\RR_2], \nonumber \eea
where $U_{1}^{(1)}$ and $U_{2}^{(1)}$ are the energies of the
nucleoside displacement with the respect to the backbone group in
each strand, $U_{3}^{(1)}$ is the energy of the H-bond stretching
in the base pairs, and $U_{4}^{(1)}$ is the energy of the
interaction of the backbone groups of different helix chains. But,
at once the term $U_{4}^{(1)}$ may be omitted because interaction
between backbone groups in a monomer link is much less than the
same interaction along macromolecule chain (the corresponding
distances differ more than twice [18]). The interactions along the
backbone chains will be taken into account in another potential
terms.

The term $U^{(2)}$ in the sum (2) describes the energy of the interaction
along the macromolecule chain. This energy term may be written in the form
corresponding to the traditional approach of the elastic rod model with
the accounting interactions between the nearest neighbors. The view of the
term $U^{(2)}$ would be concretized for definite coordinates of the
conformational rearrangements.

The term $U^{(3)}$ describes the energy of the interrelation
between the intermonomer transformations (conformational changes),
and the changes of the configuration of the helix as a whole
(deformation of the macromolecular chain). The term $U^{(3)}$ is
important under the consideration of the mesoscopic displacements
of the macromolecule structural elements, because DNA structure is
relatively soft, and under the change of the monomer configuration
the position of monomer link in a macromolecule chain changes also
[18].

The energy expression (1,2) is sufficiently complicated for the direct
use. Firstly, the vector character of the displacements which requires the
construction of the corresponding force field for the macromolecule.
Secondly, the heterogeneity of the energy terms, that makes the analytical
study practically impossible.

But, as known [18,25,27,41-47] under rearrangements of DNA
structure in the frame of the double helix the displacements of
the structural elements of the monomer link are interdependent,
and DNA structural elements move jointly along the definite
conformational pathways. Let us take this fact into account and
consider the model constructing for the basic conformational
transformations of DNA double helix: the conformational transition
of $A$--$B$ type and the helix opening.

\subsection {Conformational transitions}

Under modeling DNA conformational transitions let us introduce the
displacements of the center of masses of the pair of nucleosides
($\rr_p$) and the center of masses of the backbone groups
($\rr_q$), as well as the displacements of the nucleosides with
respect to each other in the monomer link ($\bfd$), and the same
for the backbone masses ($\bfDel$ ):
\bea \ba{llllll} \rr_p &=& {\ds \m_1\rr_1+\m_2\rr_2\over\ds m_p},\ \ &
m_p&=&\m_1+\m_2 \, ; \\[0.3cm]
\bfd &=& \rr_1-\rr_2 \, , \ \ &
\mu_p&=&\m_1\cdot \m_2 \,/\, m_p, \\
\ea
\eea
where $m_p$ is the mass of pair of the nucleosides in the monomer
link, and $\mu_p$ is the reduced mass of the nucleosides.
Accordingly, for the backbone masses we have:
\bea \ba{llllll} \rr_q &=& {\ds 1\over\ds 2}\left( \RR_1+\RR_2\right),
\ \ & m_q&=&2 \m_0 \, ; \\[0.3cm] \bfDel &=& \RR_1-\RR_2 \, , \ \ &
\mu_q&=& \m_0/2 \, . \ea \eea

In the designations (3) and (4) the expression for the n-th item of the
kinetic energy of a system takes the form:
\be K_{n}={1\over2} \left\{ m_p\dot{\rr_p}^2 +m_q\dot{\rr_q}^2 +\mu_p
\dot{\bfd}^2+ \mu_q\dot{\bfDel}^2\right\}.
\\[0.3cm]
\ee

For the further advance it is useful to put out the displacements of the
mass center of a monomer link ($\RR$) and displacements of the mass
center of a pair with respect to the backbone masses ($\rr$):
\bea \ba{llllll} \RR &=& {\ds m_p \rr_p+m_q \rr_q\over\ds M}, \ \ &
M&=&\m_1+\m_2+2\m_0 \, ; \\[0.3cm]
\rr &=& \rr_p -\rr_q \, , \ \ & m&=&m_p\cdot m_q/\, M \,. \ea \eea
Here $M$ and $m$ are the mass and reduced mass of a monomer link,
respectively. Taking into account (6) an expression for the kinetic energy
takes the following form:
\be K= {1\over2} \sum_n \left\{ M\dot{\RR}^2_{n} + m\dot{\rr}^2_{n} +\mu_p
\dot{\bfd}^2_{n} + \mu_q\dot{\bfDel}^2_{n}\right\}. \ee

Notice, the expression (7) can be simplified some more. At the
description of the dynamics of conformational transitions not all
terms of the kinetic energy are equally important.  As was
mentioned above, the potential energy associated with the last
term in the kinetic energy (7) may be omitted ($U_{4}^{(1)} \sim
0$). Thus, from the point of classical mechanics $\bfDel$ is the
cycle coordinate, and it can be ignored here. So, it is possible
to neglect the last term in the expression (7).

Let us take into account also, that in such structural transformations as
conformational transitions between the forms of $B$- and $A$- families of
a double helix, the nucleic bases in the pair displace together without
significant mutual shifts. In any case, under conformational transition
both in the initial and in the final helix forms the bases remain in a
configuration of a complementary pair [5,18,41-43,47]. So, we will assume
that under DNA conformational transitions it is true: $\bfd \approx
\dot{\bfd} \approx0$. Therefore, an expression for the model kinetic energy
for the description of conformational transitions can be written as
the following:
\be K_{tr}= {1\over2} \sum_n \left\{ M\dot{\RR}^2_{n} + m\dot{\rr}^2_{n}
\right\}. \ee

The kinetic energy in the form (8) remains still sufficiently complicated
for the modeling. And, as the last step in model construction, let us take
into account that in the conformational transformations of the
macromolecule the masses $M$ and $m$ have the primary directions of the
displacements. As it is known, under the conformational transitions of
$A$--$B$ type the mass center of DNA monomer link has the most
displacement in the plane which is orthogonal to the helix axis
[18,41-43]. Thus, for the degree of freedom of the monomer mass center it
is convenient to pass to the displacements in the plane X0Y (we put the
helix axis along 0Z, Fig.~1). In this case we have:
\beq M\dot{\RR}^2 = M\dot{\Upsilon}^2 + I \dot{\varphi}^2, \eeq
where $\Upsilon$ and $\varphi$ are the displacement and the angle of
rotation of the link in the plane X0Y, $I =I_{0Z}$ is the moment of
inertia of the monomer link with respect to a helix axis.

The necessity of taking into account the displacement on $\Upsilon$ and
torsion on $\varphi$ simultaneously is determined by the concrete
conditions of the transition. Both types of deformation are observed in
the experiment studies [5,18]. Under definite conformational transitions
these degrees of freedom change in a correlative character to minimize the
distortion of base pair stacking. The including in the model both degrees
of freedom do not cause the difficulties. For the purposes of the present
study under description of $A$--$B$ transition we will use one coordinate -
$\Upsilon$ to denote this type of the transformation.

In its turn it is also known [18,43,47,48], that under conformational
transitions the base pairs displace transversely to the helix axis, along
the axis 0X (Fig. 1b). Thus, for the component $\rr$ it is expedient to
pass to the displacement along the trajectory of the mass center motion:
$\rr\rightarrow u$.

In such a way an expression for the kinetic energy of the $A$--$B$
transition in terms of $\Upsilon$ and $u$ can be written in the scalar
form:
\be K_{tr}= {1\over2} \sum_n \left\{ M\dot{\Upsilon}^2_{n} + \m
\dot{u}^2_{n} \right\}. \ee

Under constructing the kinetic energy of the system one fact has attracted
the attention at once. After passing in the kinetic energy to the
description of the joint motions (beginning from the expression (7)) the
appeared mass coefficients don't depend on the heterogeneity of the
monomer content of DNA macromolecule. The calculated values of the masses
are shown in the Table II for A$\cdot$T and G$\cdot$C pairs.

The potential energy for the conformational transition in the selected
coordinates consists of the energy of helix bending (external component)
and the energy of base pair displacements (internal component).

The energy term $U^{(1)}$ in expression~(2) in this case comes to the
energy of the pair displacements relative to the backbone. These
displacements occur due to the changing in the sugar ring conformation
[18,41,43,47,48], and the potential energy along the coordinate $u$ corresponds
to the transition from one sugar form to another. We will describe this
energy by the function $\Phi[u]$.

The term $U^{(1)}_{3}$ in expression~(2) which describes the
energy of the $H$-bonds stretching in the base pairs has to be
omitted because the conformational transformations of $A$--$B$
type, as was mentioned above occur without significant change of
the distance between nucleosides mass centers ($\bfd \approx 0$).

The potential energy of the interaction along the macromolecule
chain (term $U^{(2)}$ in expression~(2)) includes the energy of
the bending on the component  $\Upsilon$ which is bound to have
the view akin to the bending energy of the elastic rod, and the
energy of the change of pairs stacking interactions (displacement
$u$).

Therefore, the expression for the potential energy for the description of
the dynamics of the conformational transitions of DNA macromolecule takes
the form:
\bea U_{tr} &=& {1\over2} \sum_n \biggl\{ k_{\Upsilon} \bigl[\Upsilon_{n}
- \Upsilon_{n-1}\bigr]^2 + k_{u} \bigl[u_{n}-u_{n-1}\bigr]^2
\nonumber \\[0.3cm]
&+&\Phi[u_{n}] + W \bigl[u_{n}; \Upsilon_{n \pm 1}\bigr] \biggr \}\, .
\eea
In the expression~(10) $k_{\Upsilon}$ and $k_{u}$ are the elastic
constants of the interactions along the macromolecular chain for
bending and pair displacement components, respectively.

The term with the function $W$ describes the interrelation between
the base pair displacement in $n$-th monomer and the corresponding
displacement of the $n$-th monomer with respect to $(n-1)$-th and
$(n+1)$-th monomers (the term $U^{(3)}$ in the expression (2)).
Here: $\Upsilon_{n \pm
1}$=$[\Upsilon_{n+1}-\Upsilon_{n}]+[\Upsilon_{n}-
\Upsilon_{n-1}]$=$[\Upsilon_{n+1}-\Upsilon_{n-1}]$.

The energy expression (9,10) may be used for modeling the dynamics
of DNA conformational transitions. It is important to note that
owing to passing to the description of joint motions, the mass
coefficients in kinetic energy became non-sensitive to the
nucleotide heterogeneity in DNA. Moreover, the potential energy of
the conformational transition dynamics became also non-sensitive
to DNA nucleotide content. The effect of DNA heterogeneity may
reveal only in some sensitivity of model elastic constants ($k$)
to the nucleotide sequence in macromolecule because of known
sensitivity to the sequence of the potential energy of
interactions along the DNA chain [18]. These properties of the
model are in common agreement with the known data about
non-sensitivity of DNA $A$--$B$ transitions to the nucleotide
content [41,48].

\subsection {Opening}

Other structural transformation that has attracted considerable
interest in DNA physics is the pair opening in DNA double helix.
This process offers the conformational transformation, as a result
of which the hydrogen bonds in the complementary pairs disrupt,
and the nucleic bases turn out from the double helix [18].

It is considered [27,46] that the most probable trajectory for
double helix opening is the turning of bases of the pair around
backbone chains to the sides of the helix grooves (so called
"opening", the direction of rotation on the angles $\theta_1$ and
$\theta_2$, Fig. 1b). The other possible pathway connected with
the stretching of base pair along the H-bonds (OY axis on Fig. 1b,
"stretching") is assumed as less probable because of known strong
rigidity of the backbone chains in the direction of OY axis [42].
The "stretching" process can realize under the temperature
increase in the conditions of melting. The modeling of the base
pair stretching in double helix was made in a number of works
[30,49-52]) with the help of one-component model. In the present
study the "opening" trajectory will be considered, as the most
probable in the natural conditions where the mesoscopic
transformations occur.

Let us consider the form of kinetic energy for description of the
double helix opening. In the four-mass model (1,2) we will
introduce the displacements of mass centers of the nucleosides
with respect to the backbone, and the displacements of mass
centers of the nucleotides as a whole. For the displacements of
masses in the n-th link of the $i$-th DNA strand we will write the
following equations:
\bea \ba{llllll}
\bfwp_i &=& {\ds \m_i{\rr}_i + \m_0\RR_i\over\ds \M_i}, \ \
&\M_i&=&\m_i+\m_0 \, ; \\[0.3cm]
\bfro_i &=& \RR_i-{\rr}_i, \ \
& \mu_i&=&\m_i\cdot \m_0/\, \M_i \, , \ea
\eea
where $\M_i$ and $\mu_i$ are total and reduced masses of the nucleotide in
the $i$-th strand.

Since the main process in the opening transformation is the
extension the H-bonds in the base pair, it may be assumed that
under helix opening the nucleosides remain rigid. We will consider
that in DNA opening the mobility of the nucleosides is brought to
the turnings as a whole around the backbone. Thus:
\be \dot{\bfro_i}^2 \longrightarrow {l_i}^2 \dot{\theta_i}^2,
\ee
where $l_i$ is the reduced length of the pendulum-nucleoside and
$\theta_i$ is the angle of its deviation from the equilibrium state (Fig.
1b).

Under the conditions (12) the kinetic energy of a system takes the form:
\be K={1\over2} \sum_{i,n} \left[ \M_i\dot{\bfwp}^2_{i,n} + \mu_i {l_i}^2
\dot{\theta}^2_{i,n} \right]. \ee

The expression (13) remains still complicated for the modeling of
the DNA opening. Let us pass to the consideration of the joint
motions of the structural elements in the double helix link. With
this purpose we will single out the displacements of the mass
center of both nucleotides (mass center of monomer link) and the
relative displacements of nucleotides in the link. These
displacements and the corresponding expressions for the masses
have the form:
\bea \ba{llllll} \RR &=& {\ds \M_1\bfwp_1 + \M_2\bfwp_2\over\ds M},
\ \ &M&=& \M_1 + \M_2 \,; \\[0.3cm] \bfxi &=& \bfwp_1-\bfwp_2, \ \ &
\mu&=& \M_1\cdot \M_2 \, /\, M \,. \ea \eea

Let us introduce also the joint turnings of the nucleosides:
\bea
\ba{llllll}
\Omega &=& {\ds J_1\theta_1-J_2\theta_2\over\ds J}, \ \ &
J&=&J_1+J_2 \ \ (J_i=\mu_il_i^2)\, ; \\[0.3cm]
\sigma &=& \theta_1+\theta_2, \ \ & j&=&J_1\cdot J_2 \, / \,J \,.
\ea \eea
Here the parameters $J$ and $j$ have a meaning of the joint and
relative inertia moments of the nucleosides in the pair. At
writing (15) it was taken into account, that the angles $\theta_1$
and $\theta_2$ are disdirective (Fig. 1b). The angle $\sigma$
correspond to the known parameter "opening" in the classification
of DNA transformations [53].

Taking into account the expressions (14) and (15) the kinetic energy of
the opening can be presented as:
\be K_{op}={1\over2} \sum_n\left\{ M\dot{\RR}^2_{n} +\mu \dot{\bfxi}^2_{n}
+J\dot{\Omega}^2_{n} +j\dot{\sigma}^2_{n} \right\}. \ee

The values of the parameters for the kinetic energy (16) are
calculated and adduced in the Table III. It is significant that
these parameters are practically independent from the kind of DNA
pair. Thus, as in the case of the description of conformational
transitions, both for helix opening the kinetic energy of the
heterogeneous macromolecule assumes the homogeneous form after
passing to the joint motions of the structural elements.

Then, for further simplification we will take into account that in
the opening process the motions of the structural elements take
place along the definite pathway. We will presume the uniform
distribution of energy among the moved nucleosides, and in the
wake of [27,46] that the opening occur in such way that:
$\theta_1\approx \theta_2$ and $|\Omega| \ll|\sigma|$. So, we will
suppose that: $\Omega \sim \dot{\Omega} \sim 0$.

It is also known that when the base pair opens to the side of the
helix groove, the monomer link as a whole displaces, and the
macromolecule bends [44-46]. On this way the nucleotides in the
pair displace jointly, that is corresponded to: $\bfwp_1\approx
\bfwp_2$ , $|\bfxi| \ll|\RR|$, and $\RR\rightarrow Y$, where $Y$
is the coordinate of the displacement to the side of the helix
grooves (Fig. 1b).

Thus, an expression for kinetic energy of the double helix opening
transformations acquire the two-component form:
\be K_{op}={1\over2}\sum_n\left[ M\dot{Y}^2_{n} +j\dot{\sigma}^2_{n}
\right]. \ee

The expression for potential energy for the opening pathway may be
presented in the form which looks like the expression (10):
\bea U_{op} &=& {1\over2} \sum_n \biggl\{k_{Y}\bigl[Y_{n}-Y_{n-1}\bigr]^2
+ k_{\sigma} \bigl[\sigma_{n}-\sigma_{n-1}\bigr]^2
\nonumber \\[0.3cm]
&+& \Phi[\sigma_{n}] + W\bigl[\sigma_{n}; Y_{n \pm 1}\bigr] \biggr \}\, .
\eea
In the expression (18) $k_{Y}$ and $k_{\sigma}$ are the elastic
constants of the interaction along the chain, $\Phi[\sigma]$ - the
potential energy of the pair opening which also includes the nucleoside
turning around the backbone chains, and the term $W$ describes the
energy of the interrelation between the components $\sigma$ and $Y$.

The shape of the potential $\Phi[\sigma]$ must allow for the
description of transition from the ground state of the double
helix to the metastable state. Considering the form of the
potential $\Phi[\sigma]$ it must be taken into account that on the
pathway of opening the interaction of the opened bases with
surrounded water have lead to the formation a set of metastable
states of the base pair with molecules of water. Such states
(preopened) have been recently found in the quantum-mechanical
calculations [54,55], and observed in the experiment [13]. These
results show that the double helix opening is the stage-by-stage
process, and the potential $\Phi[\sigma]$ has bistable (or more
correctly, multistable) form. In the case of the description of
the transition from a close state of the double helix to the
preopened one the potential $\Phi[\sigma]$ must have the shape of
double-well function.

When modeling the potential energy of DNA opening it is important
to allow the dependence of the energy of the pair opening on the
kind of a pair (the difference in hydrogen bonding of A$\cdot$T
and G$\cdot$C). In other words, in the case of opening the energy
$\Phi[\sigma]$ is sensitive to the base pair content in DNA.

\section {UNIFYING MODEL FOR DNA INTERNAL MOBILITY ON THE
MESOSCOPIC SCALES}

As it is seen from the results of the previous section of this
work, the relatively simple two-component models can be
constructed for description of various DNA transformations in the
frame of double-helical state. The models for different DNA
mesoscopic rearrangements are very similar. They include the
external component which is the degree of freedom of the elastic
rod, and the internal component which is the characteristic
coordinate of structural transformation. This is the basic
construction that may be modificate by including the additional
terms - components of both types. Let us write the expression for
the energy of DNA mesoscopic transformation in the unifying form:
\bea E & = & {1\over2} \sum_n \biggl\{ M \dot{R}^2_{n} + k_{1} \bigl[R_{n}
- R_{n-1}\bigr]^2
\nonumber \\[0.3cm]
& + & m \dot{r}^2_{n} + k_{2}\bigl[r_{n}-r_{n-1}\bigr]^2 +\Phi[r_{n}] +
W\bigl[ r_{n}; R_{n \pm 1}\bigr] \biggr\}\, . \eea
In the expression (19) $R$ and $r$ are the degrees of freedom of external
and internal components, respectively, $M$ and $m$ -- the full and reduced
monomer masses, $k_{1}$ and $k_{2}$ -- the elastic constants of the
interactions along the macromolecular chain for external and internal
components.

The function $\Phi[r]$ describes the potential energy of the
internal subsystem on the pathway of a conformational
transformation from one stable state to another. Hence, it must
have the shape of the double well (Fig.2). The form of the double
well is usually known from the experimental data. For the
conditions when one of the helix form is more stable (for example
at the physiological conditions the most stable is the B-form of
the double helix [18]) the function $\Phi[r]$ must be taken in the
non-symmetric view with two non-equivalent stable states: ground
and metastable (Fig. 2a). That is true both for the conformational
transitions such as $A$--$B$, and for the transition in the
preopened state. If the conformational transition takes place
under equilibrium of the states (for example in the interval of
$A$--$B$ equilibrium), the function $\Phi[r]$ must have the
symmetric view with two equivalent states (Fig. 2b).

The term $W$ reflects the energy of the interrelation between the
components, and may be written in the form:
\bea W = \chi F[r_{n}] \biggl\{R_{n+1}-R_{n-1}\biggr\} \, .\eea
Here $\chi$ is the coefficient, the potential function $F(r)$
describes the change in the interrelation energy under the
transition of the internal component from one stable state to
another. By its physical meaning the potential $F(r)$ is the
function that increases from the ground state of the system and
decreases near the another stable state. In the case of the
description of the transition between the equilibrium states this
function must have the symmetric form in accordance to physical
equivalence of the states. A sample view of this potential
function is shown in Fig.~2a and 2b..

The two-component models of type (20) were used for modeling the
transformations of bistable chains of different nature. Firstly
such model was used  under study of the dynamical properties of
the systems with H-bonds [56,57], where the bistable chains with
equivalent stable states were considered. The two-component model
of bistable macromolecule with non-equivalent stable states was
introduced for the study of the nonlinear dynamics of local
$A$--$B$ transitions in DNA [58-61]. The same models were used for
the description of DNA melting [62], and the nonlinear dynamics of
the endothermic transitions [63].

Let us pass to the continuum approximation, that is frequently used in
analytic studies of models. In this case it is supposed that
conformational excitation embrace some piece of the macromolecule chain
which is much larger than the chain step. In the continuum approximation
the expression for the energy (19) may be written in the form:
\bea E &=& \int {dz\over2h} \biggl[ M\left( \dot{R}^2 + s_{1}^2{R'}^2
\right)+m\left( \dot{r}^2 +s_{2}^2{r'}^2 \right) \nonumber
\\[0.3cm]
&+& \Phi(r) + 2\chi h F(r) R' \biggr]\, , \eea
where the expression for the interrelation energy (20) was also taken into
account. In the expression (21) $R=R(z,t)$ and $r=r(z,t)$, $R'$ and $r'$
are the derivations on z, $s_{1}^2=k_{1} h^2 / M$ and $s_{2}^2=k_{2} h^2 /
m$.

Notice, when modeling DNA conformational transformations the
sensitivity of the elastic constants to the nucleotide sequence
was not considered. Really, the elastic constant $k_{2}$ (and to
some extent $k_{1}$) reflect the value of stacking interactions
between the neighbor pairs in the helix. In its turn the stacking
depends on the kind of the interacting pairs (see Refs. [18,27,48]
and the references therein). This sensitivity of the elastic
constants has an effect on the form of the fine structure of
conformational excitation, and can be the object of special
consideration under study of such effects on the quantitative
level.

The equations of motion for two-component excitation with the energy (21)
have the form:
\bea \ddot{R} & = & s_1^2 R'' +\chi_{1} \frac{dF}{dr}r' \, ;
\\
\ddot{r} & = & s_2^2 r'' - \frac{1}{2m} \frac{d\Phi }{dr}- \chi_{2}
\frac{dF}{dr} R'\, , \eea
where $\chi_{1}= \chi h / M$ and $\chi_2=\chi h / m$ - the constants of
the interrelations.

The equations (22,23) describe the dynamics of the two-component lattice.
It is very much the same that the dynamics of one-dimensional molecular
crystal [64]. The significant distinction is the including the terms with
coefficient $\chi$. As was mentioned above, the interrelation between the
components for such "soft" lattice as DNA macromolecule have been taken
into account because of direct evidences of the experiment.

For better understanding the role of interrelation let us consider the
dynamics of two-component model (21) in the neighborhood of it ground
state. For the small deviation from the ground state: $r=r_{0}+\xi$, $\xi
\ll r_{0}$ and $R\sim\rho \ll r_{0}$ the expressions for the potential
functions may be written as: $\Phi \approx a_{0}\xi^2$ and $F \approx
b_{0}\xi $, where $a_{0}={1\over2}{d\Phi^2\over dr^2} |_{r_0}$, and
$b_{0}={dF\over dr}|_{r_0}$ are the first nonzero terms in the potential
expansions on $\xi$.

The equations of motion (22,23) for the small deviations from the ground
state have the form:
\bea \ddot{\rho} & = & s_1^2 {\rho}'' +\chi_1 b_{0} {\xi}' \,;
\\
\ddot{\xi} & = & s_2^2 {\xi}'' - \frac{1}{m} a_{0}{\xi}- \chi_2 b_{0}
{\rho}'\ . \eea

We will find the decisions of the equations (24,25) in the form of
oscillating waves: $\xi=\xi_{0} \exp {i(wt- \k z)}$ and $\rho=\rho_{0}
\exp {i(wt- \k z)}$. After the wave substitution the equations (24,25)
lead to the dispersion equation which may be written as:
\be (\omega^{2} - s_{1}^{2}\k^{2})(\omega^{2} - s_{2}^{2}\k^{2} -2a_{0} /
m) = \k^{2}b_{0}^{2}\chi _{1} \chi _{2} \, . \ee

For very small wave vector $\k$ ($\k^{2}\sim 0$) from (26) the two limit
frequencies follow: $\omega^{2}\approx\omega_{0}^{2} = 2a_{0} / m$ and
$\omega^{2}\approx 0$. Thus, in the two-component system (21) we have two
type of phonon vibrations. One, with the limit frequency $\omega_{0}=
{1\over m}{d \Phi^{2}\over d r^{2}}|_{r_{0}}$ - optical branch, that
determines the dynamics of the internal component. Second type, with the
$\omega\rightarrow 0$ under $\k\rightarrow 0$, is the acoustic branch.

The optical branch correspond to the definite optical vibration of
the double helix, considered in the linear four-mass model
[38-40]. That may be used for more precise selection of the actual
degrees of freedom for modeling the mesoscopic mobility, and the
evaluation of the parameters of the potential functions.

For small $\k$ ($\k^{3}$ and $\k^{4}$ $\sim 0$, the quadratic
approximation) the expression (26) is equivalent to the following:
\be (\omega^{2} - c_{1}^{2}\k^{2})(\omega^{2} - c_{2}^{2}\k^{2} -
\omega_{0}^{2}) = 0 \, ,\ee
where $c_{1}^{2}= s_{1}^{2}-s_{\ast}^{2} $ and $c_{2}^{2}=
s_{2}^{2}+s_{\ast}^{2}$, $s_{\ast}^{2}=b_{0}^{2}\chi _{1} \chi
_{2}/\omega_{0}^{2}$.

So, in the quadratic approximation for the dispersion law the
analyzed system is the same that is true for usual two-component
lattice [64] but with the force constants $c_{1}$ and $c_{2}$. The
constants include the contribution of the energy of interrelation
($s_{\ast}$). As seen from their values the interaction of the
sublattices leads to the softening of the external sublattice -
the degrees of freedom of the elastic rod, and to the greater
rigidity of the internal sublattice. Notice that the effect of the
component interrelation is larger than the frequency of the
corresponding conformational vibration is lower. As known (see
[38-40] and reference therein), the frequency of DNA
conformational vibrations are sufficiently low ($<$100{cm}$^{-1}$)
So, the effect of the component interrelation is important, and
must be taken into account under the modeling DNA mesoscopic
transformations.

\section {STATIC STATES OF THE TWO-COMPONENT MODEL}

In this section we will consider the static states of the unifying
model (21), which may be compared with experiment to demonstrate
the correctness of the approach. To the static states we will
assign the stable static states of the system and the static
excitations which can be stable under some conditions. The stable
static excitations of the system may be observed as static
deformations of the real macromolecule. So, let us determine the
static states of the two-component model (21).

For more general character we will study the static states of the
system without definition of concrete expressions for the
potential functions. In accordance with the shape of the
potentials we will assumed that the potential energy of the system
will have three extremums: two minimums ($r_{0}, r_{2}$) and one
maximum ($r_{1}$) between them. In the ground state $r=r_0$
(B-form for DNA) we will consider $\Phi(r_0)= F(r_0)= 0$. In the
metastable state ($r=r_2$) $\Phi(r_2)\not=0$, $F(r_2)\not=0$ (Fig.
2a). For the conditions when the macromolecule have two equivalent
stable states (Fig. 2b) the function $F(r)$ for $r_{2}$ give
$F(r_2)=0$.

For providing of the bistabile form of the model energy it is
necessary to satisfy the definite conditions for the extremum
points [65]:
\be \Phi(r_{1,2}) \geq {\chi^{2}\over k_{1}}\,F^{2}(r_{1,2}) \ ,\ee
that is an inequality for $r_1$ state always, and for $r_2$ state
is inequality in the case of the non-equivalent states and -
equality for the case of equivalent states.

\subsection{Static excitation in the macromolecule with metastable
states}

Let us find the static excitations of the macromolecule with
metastable states in the monomer link. We will assume: $r=r(z)$
and $R=R(z)$, and $R'(z)=\tau/h $. Here the value $\tau(z)$ is
coined for the determination of the macromolecule deformation on
the external component (torsion or bending, in dependence of
concrete model). The equations described the static excitations
follow from the Eqs. (22,23) of the system:
\bea
 R'' &+& {\ae_{1}} \frac{dF}{dr}r' = 0 \, ;
\\
r'' &-& \frac {1}{2\kappa} \frac{d\Phi }{dr}- {\ae_{2} } \frac{dF}{dr} R'
= 0 \, . \eea
In the Eqs. (28,29) $\ae_{1}=\chi/k_{1}h$, $\ae_{2}=\chi/k_{2}h$ and
$\kappa=k_{2}h^{2}$.

After one time integration of the Eq. (29) we obtain:
\be
 R' + \ae_{1}\, F(r) = 0 \, ,
\ee
where the constant of integration put to zero according to the initial
conditions for the stable state. The deformation of the external component
of the macromolecule chain may be written from the Eq. (31) as:
\be \tau = - {\chi\over k_{1}} F(r)\, . \ee
As seen from expression (32) the deformation of the chain in the model is
proportional to the function $F(r)$.

After substitution the expression for $R'$ in the Eq. (30) and one time
integration we obtain the equation for the internal component:
\bea r'^{2} - \frac {1}{\kappa} \Phi(r) + \ae_{1} \ae_{2} F^{2}(r) = C \,,
\nonumber \eea
which may be rewritten in the form:
\be
 r'^{2} + Q (r) = 0 \, .
\ee

The equation (33) have the view of the energy of the mechanical system
with the potential energy: $Q(r)= - \frac {1}{\kappa} \Phi(r) + \ae_{1}
\ae_{2} F^{2}(r) - C$.

The solution of the Eq. (33) may be found as the conversion of the
integral
\be
 \int\limits_{r(0)}^{r(z)} \, {du\over \sqrt{-Q(u)}} = z.
\ee

The view of the effective potential $Q(r)$ is determined by the concrete
form of the potential functions $\Phi$ and $F$. When fulfilling the
condition (28) the effective potential $Q(r)$ have the form of a double
hump (see Fig.~3a).

Let us consider the possible solutions of the equation (34). We
will find the excitations with a restricted trajectory and the
asymptote of one of the stable states of the system because such
excitations are the most stable in the real conditions. For the
potential $Q(r)$ with two non-equivalent humps we will search the
excitation with the asymptote of the metastable state $r_2$ (Fig.
3a). In this case it is convenient to present the polynom $Q(r)$
in the form:
\be
 Q (r) = - Q^{2}_{m} (r_2-r)^{2}(r- \eta_1)(r-\eta_2).
\ee
In the expression (35) $Q^{2}_m$ is the combination of the model
parameters, $r_2$, $\eta_1$ and $\eta_2$ ($\eta_1 < \eta_2 < r_2$)
are the zeros of the polynom $Q(r)$ with the meaning $C= - \frac
{1}{\kappa} \Phi(r_2) + \ae_{1} \ae_{2} F^{2}(r_2)$ for the
boundary conditions of the metastable state ($r=r_2$ and
$r'_2=0$). We will interest in the excitations with the trajectory
restricted by the interval: $\eta_2 \leq r \leq r_2$ (see Fig.
3a).

For the polynom $Q(r)$ in the form (35) the integral equation for the
function $r(z)$ have the form:
\be
\int\limits_{\eta_2}^{r(z)} \, {du\over
(r_2-u)\sqrt{(u-\eta_1)(u-\eta_2)}} = \pm Q_{m}z. \ee

The integral~(36) have the table view. After integration we obtain the
following relation:
\bea && ln \left\{{1\over (r_2-r)(\eta_2-\eta_1)} \left[
\sqrt{(r_2-\eta_1)(r-\eta_2)}
\right.\right. \nonumber \\[0.3cm]
 &\!\!\!-\!\!\!&  \left.\left. \sqrt{(r_2-\eta_2)(r-\eta_1)}
\right]^2 \right\}= \pm Q_{m}z \sqrt{(r_2 -\eta_2)(r_2-\eta_1)}\, .
\nonumber \eea
Or, after rewriting:
\bea && \exp (q_{m}z) \left(r_2-r\right)\left(\eta_2-\eta_1\right)
\nonumber \\[0.3cm]
 &\!\!\!=\!\!\!&  \left[ \sqrt{(r_2-\eta_1)(r-\eta_2)}-
 \sqrt{(r_2-\eta_2)(r-\eta_1)}\, \right]^{2}\, ,
\eea
where $q_{m}= \pm Q_{m}\sqrt{(r_2-\eta_2)(r_2-\eta_1)}$.

For convenience let us designate: $r_a=r_2-\eta_2$ and
$r_b=\eta_2-\eta_1$. The values $r_{a}$ and $r_{b}$  reflect the
form of the potential $Q(r)$ (Fig. 3a). Their substitution
to the Eq. (37) lead to the expression:
\bea && \left(r_{2}-r\right)\left[r_b\,\exp (q_{m}) +2r_a +r_b \right]-
2r_a(r_a+r_b) \nonumber \\ [0.3cm] &\!\!\!=\!\!\!&
2\sqrt{r_a(r_a+r_b)\left[(r_2-r)^2 -(r_2-r)(2r_a+r_b)
+r_a(r_a+r_b)\right]}\, . \eea

After raising to the second power of the whole expression (38) it is easy
to obtain the form of the internal component of the static excitation:
\bea r(z) &\!\!\! = \!\!\!& r_2-r_{\rm ex}(z) \, , \\ [0.3cm] r_{\rm ex}
(z) &\!\!\! = \!\!\!& {2r_a(r_a +r_b)\over r_b\ch (q_{m} z) +2r_a+r_b} \,
. \nonumber
\eea

As seen, at $z\rightarrow 0$ (the middle point of the excitation) $r_{\rm
ex}\rightarrow r_a$ and $r(z)\rightarrow \eta_2$. At $z\rightarrow
\infty$, $r_{\rm ex}\rightarrow 0$ and $r(z)\rightarrow r_2$ (Fig. 3b).

The external component of the excitation according to the Eq. (31) have
the form:
\be
\tau_{\rm ex}(z) = -{\chi\over k_{1}} F\left[ r_2-r_{\rm ex}(z)\right].
\ee
At $z\rightarrow \infty$, $r_{\rm ex} \rightarrow 0$, $\tau_{\rm ex}
\rightarrow \tau_2$ and at $z\rightarrow 0$, $r_{\rm ex} \rightarrow r_a$,
$\tau_{\rm ex} \rightarrow \tau_a = -{\chi\over k_{1}} F[r_2-r_a]$, where
$0>\tau_a>\tau_2$.

Thus, as seen from the Eqs. (39,40) the static excitation has the
form of the bell with the width that is proportional to
$q^{-1}_{m}\sim (r_{a}^{2}+r_{a}r_{b})^{-{1\over 2}}$ and the
height -- $r_{a}$. At the middle point of the excitation the
components have the values that are close to the same for the
ground state (Fig. 3b). On the edges of the excitation the
components have the values of the metastable state.

The obtained results may be used for comparison with the observed
static deformation of the macromolecule. The set of data on the
static conformational deformation in DNA was published recently.
So, in the work [66] it was found the structure of DNA-protein
complex.  The reconstruction of DNA structure in this complex
[5,67] show that protein has induced $B \rightarrow A$ transition
in DNA, and the DNA fragment is bent and looked as the bell.
Besides that, the conformation of the double helix in the center
of deformation is close to ground state ($B$-form), and on the
edges - to $A$-form.

Notice, that the static excitation (39,40) is commonly agree with
the observed form of DNA deformation. As in experiment [5,66,67]
the obtained static excitation has the maximum deformation on the
edges of the fragment ($\tau=\tau_2$), and the minimum - at the
central part ($\tau=\tau_a$, which $\tau_2<\tau_a<0$) (Fig. 3b).
It is also important, that the theory gives for the internal
component the conformation of the metastable state on the edges of
the excited fragment, and the conformation closed to the ground
state for the central part. That is also in accordance with the
experiment [5,66,67].

\subsection{Static excitation in the macromolecule with equivalent
states in the monomer link}

Let us consider the static excitations of the two-component chain (21)
with equivalent states in the monomer link. As in previous case the form
of the excitation for internal and external components is determined by
the Eqs. (32-34). For the equivalent states in the monomer the effective
potential $Q(r)$ in the Eq. (33) has the shape of the double hump with
equal humps (Fig. 4a). We will find the solution of the Eq. (34) for the
boundary conditions of the stable states ($r \rightarrow r_0$ or $r_2$ and
$r'\rightarrow 0 $ when $z\rightarrow \infty$). For these conditions the
constant $C$ in the polynom $Q(r)$ is equal to $0$. It is convenient to
present the potential $Q(r)$ in the form:
\be
Q (r) = - Q^{2}_{e} (r_2-r)^{2}(r-r_0)^{2}, \ee
where $Q^{2}_{e}$ is the combination of the model parameters.

The substitution of the expression (41) into the Eq. (34) and
integration gives the relation:
\be {1\over (r_2-r_0)} \, ln \left| \,{r_2 - r(z)\over r_0 - r(z)} \,
\right| = \pm Q_e z \, . \ee
After rearrangement of the Eq. (42) we obtain the expression for the
internal component of the excitation in the following form:
\be
r(z) = r_{e} + r_{d} {\rm th} (q_{e}z),
\ee
where $r_e=(r_2 + r_0)/2$ ($r_e=r_1$ for symmetric bistable potential
$\Phi$), $r_{d}=(r_2-r_0)/2$ and $q_e = \pm r_d Q_e$.

As seen from Eq. (43) at $z\rightarrow \pm \infty$, $r(z)
\rightarrow r_2$ or $r_0$ in dependence on the sing of $q_e$. At
$z=0$, $r(z)=r_e$ (Fig. 4b). The width of the excitation is
proportional to $(r_{d})^{-1}$.

For external component in accordance with Eq. (32) and the form of
the potential function $F(r)$ in the case of equivalent states
(Fig. 2b) we have the following: when $z\rightarrow \pm \infty$
then $\tau \rightarrow 0$, and $\tau(0) = \tau_e= -{\chi\over
k_{1}} F[r_e]$ at $z=0$ (Fig. 4b). Thus the external component has
another form in compare with the case of non-equivalent states. In
the central part of the macromolecule fragment the deformation
have the maximum value, and on the edges of the excitations the
deformation is absent.

These results may be compared with the data of molecular images of
the protein-induced $A$--$B$ transformation of DNA [5]. In the
work [5] the structure of the crystal DNA-protein complex [68] was
reconstructed in all-atom presentations for DNA fragment. As seen
from the data of [5], the DNA portion in the complex is in the
$B$-form, but free portion of DNA remains in the $A$-form. At the
$A$--$B$ junction DNA macromolecule bent [5]. The observed DNA
transformation is in agreement with the results of the present
modeling. As in experiment, the static excitation for the DNA with
equivalent states is caused by the transition between the forms of
the double helix, and has the maximum of macromolecule deformation
at the center of $A$--$B$ junction (Fig. 4b).

The obtained results have also a good accordance with the data on
intrinsically bent DNA [10,11]. As considered, the DNA fragments
with the homogenic sequence of A$\cdot$T pair (A-tract) cause the
bistability in DNA bending [10-12]. In studied DNA the portion
with A-tract alter with the DNA portions with the averaged
sequence of the base pairs. As supposed, the A-tract have two
possible conformations: one conformation that is the same as for
whole macromolecule, and another conformation that differs from
the remainder part of double helix. Under some conditions the
A-tracts transit to another form, and this is considered as the
reason of DNA bending [10].

The observed DNA bending [10,11] may be explained with the help of
the results of the present modeling also. According to the case of
two equivalent states in the monomer link the bending of DNA
fragment is the result of the transition from one state to
another.

It is essential that the static excitation in the internal
component (43) provides the external component bending to one side
for both type (there and back) of transitions (Fig. 4b).
Therefore, the alternating of the forms in the internal component
leads to the bending to the side of one groove of the
macromolecule. If the length of the excitation is divisible to the
helix step then the bending of the macromolecule as a whole would
realize in one side and in one plane. Such mechanism of the
intrinsically bending of DNA is in an agreement with the observed
properties of the bent macromolecule [10-12,69].

\section{CONCLUSIONS}

The performed study shows the possibility of constructing the
relatively simple, two-component models for the description of DNA
structure transformations on the mesoscopic scales. As was seen for
a number of conformational transformations in the frame of the
double-helical state the model may be formulated in unifying form.
It was shown that on the pathways of the mesoscopic
transformations the DNA macromolecule behaves as homogeneous
system in the heterogeneous potential field.

The comparison of the static excitations obtained in the constructed
models with the observed deformations of the DNA fragments show the
qualitative agreement between the theory and experiment. From this
correlation the conclusion follows that the observed DNA
deformations occasioned by intrinsical localized excitations -- static
conformational solitons by their nature. The obtained results may
be considered as the direct evidence of the localized
excitations existence in DNA macromolecule.

\section {ACKNOWLEDGMENTS}

Author would like to to express his thanks to A.G. Naumovets for
helpful discussion, to R. Lavery and J. Maddocks for the possibility
to present the results of this work on the Workshop "DNA and beyond:
Structure, Dynamics and Interactions", and to W. Olson for providing
the reprints of her publications.

\vspace{\baselineskip}
\vspace{\baselineskip}

\begin{center}

TABLE I. The parameters of the four-mass model
\vspace{\baselineskip}

\begin{tabular}{||c||c|c|c|c||}
\hline Nucleo-&$\m_i^{\star}$&$\M_i$&$l_0$ & $\Theta_0$ \cr
side&($a.u.m.$)&($a.u.m.$)&(\AA)& (degr.) \cr \hline &&&& \cr Ade
&203(134)&312&5.1&25 \cr Thy &194(125)&303&4.6&35 \cr Gua
&219(150)&328&5.4&23 \cr Cyt &179(110)&288&4.5&30 \cr &&&& \cr \hline
\end{tabular}
\vspace{\baselineskip}

$^{\star}$ In the bracket there are the masses of the nucleic bases.
\end{center}
\vspace{\baselineskip}

\begin{center}
TABLE II. The model parameters for conformational
transition in DNA \vspace{\baselineskip}

\begin{tabular}{||c||c|c|c|c|c||}\hline
Base &$M$ &$m$ &$\mu_p$ &$\mu_q$ &$I_{0Z}$ \cr
pair&($a.u.m.$)&($a.u.m.$)&($a.u.m.$)&($a.u.m.$)&
($10^{-37}$g$\cdot$cm$^2$) \cr \hline &&&&& \cr
A$\cdot$T&615&140.7&99.2&54.5&33.7 \cr
G$\cdot$C&616&140.8&98.5&54.5&34.5 \cr &&&&& \cr \hline
\end{tabular}
\end{center}
\vspace{\baselineskip}

\begin{center}
TABLE III. The model parameters for DNA opening \vspace{\baselineskip}

\begin{tabular}{||c||c|c|c|c||}\hline
Base & $M$& $\mu$& $J$& $j$ \cr pair&($a.u.m.$)& ($a.u.m.$)
&($10^{-37}$g$\cdot$cm$^2$) &($10^{-37}$g$\cdot$cm$^2$) \cr \hline &&&&\cr
A$\cdot$T&615&153.7&5.51&1.36 \cr G$\cdot$C&616&153.4&5.80&1.38 \cr &&&&
\cr \hline
\end{tabular}
\end{center}

\vspace{\baselineskip}
\vspace{\baselineskip}

FIGURE CAPTIONS
\vspace{\baselineskip}

Fig. 1. The four-mass model
for the DNA monomer link: \, (a) -- the fragment of the
polynucleotide chain and a pendulum-nucleoside construction, the
nucleoside is shown by the dot-dash line, $\ast$ - the center of
masses of nucleoside; (b) -- the masses displacements in the plane
of the complementary pair; (c) -- the double chain of the
nucleosides and backbone groups masses.

\vspace{\baselineskip}

Fig. 2. A sample view of the potential functions $\Phi(r)$ and $F(r)$:
 \, (a) -- the case of the double well with non-equivalent stable states;
(b) -- the case of the equivalent states.

\vspace{\baselineskip}
Fig. 3. Static excitation in the case of non-equivalent stable
states for DNA monomer link: (a) -- the view of the effective
potential; (b) -- the form of the excitations for internal and
external components.

\vspace{\baselineskip}
Fig. 4. Static excitation in the case of two equivalent states for DNA
monomer link: (a) -- the view of the effective potential; (b) -- the form
of the excitations for internal and external components.

\end {document}
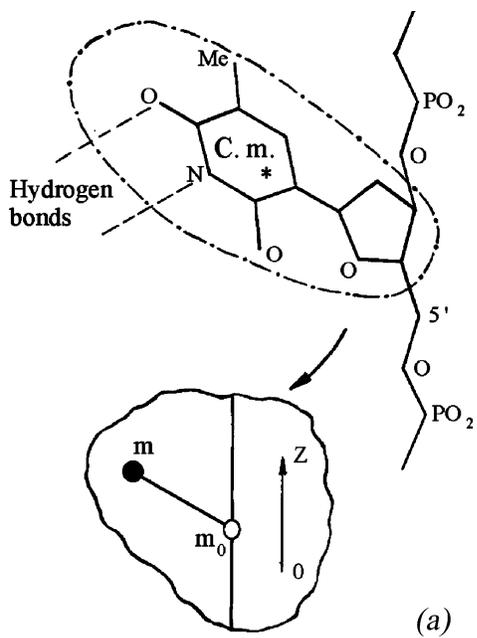
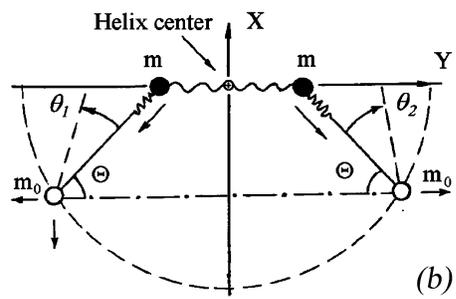
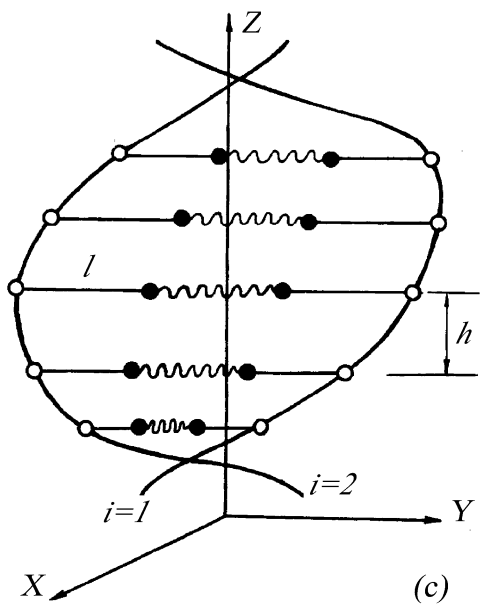
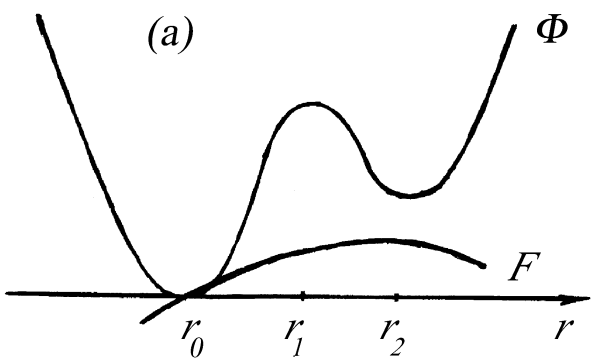
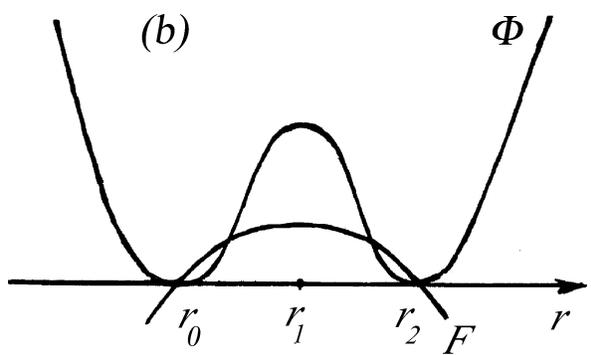
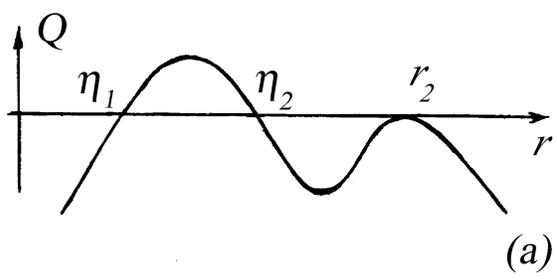
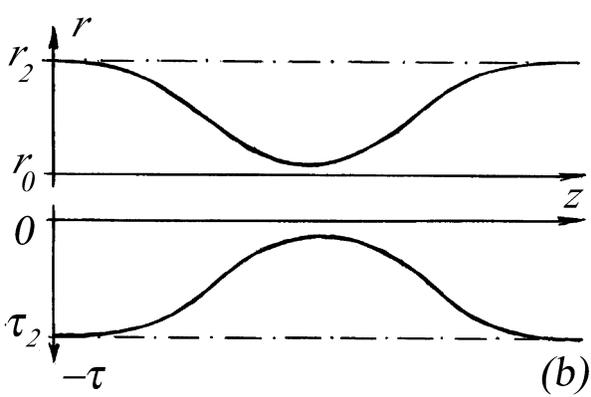
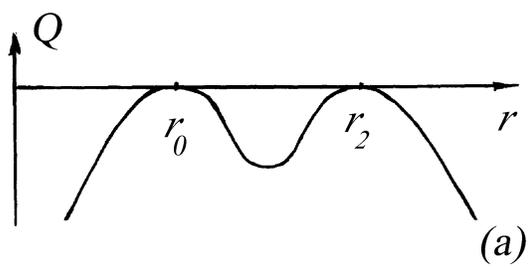
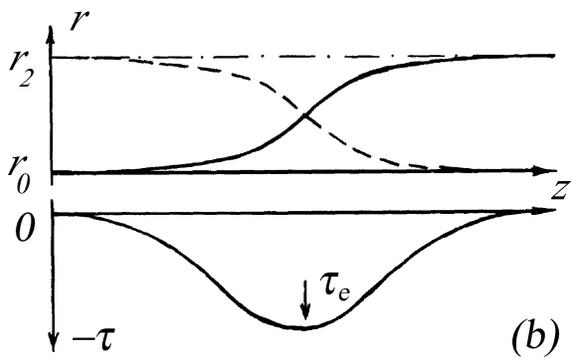
